\documentclass[%
 reprint,
 amsmath,amssymb,
 aps,
floatfix,
]{revtex4-2}

\usepackage{graphicx}% Include figure files
\usepackage{dcolumn}% Align table columns on decimal point
\usepackage{bm}% bold math
\begin{document}

%\preprint{APS/123-QED}

\title{Polarization switching  of quasi-trapped modes and near field enhancement in bianisotropic all-dielectric metasurfaces }% Force line breaks with \\
%\thanks{A footnote to the article title}%

\author{Andrey B. Evlyukhin$^{1}$}
\email{a.b.evlyukhin@daad-alumni.de}
\author{Maria Poleva$^{2}$}
\author{Alexey Prokhorov$^{3,4}$}
\author{Kseniia Baryshnikova$^{2}$}
\author{Andrey E. Miroshnichenko$^{5}$}
\author{Boris N. Chichkov$^{1}$}
\affiliation{$^1$Institute of Quantum Optics,  Leibniz Universit{\"a}t Hannover, Welfengarten Street 1, 30167 Hannover,  Germany}
\affiliation{$^2$School of Physics and Engineering, ITMO University, Saint Petersburg, 197101, Russia}
\affiliation{$^3$Moscow Institute of Physics and Technology,
9 Institutsky Lane, Dolgoprudny 141700, Russia}
\affiliation{$^4$Department of Physics and Applied Mathematics,
Vladimir State University named after Alexander and Nikolay Stoletovs,
Vladimir 600000, Russia}
\affiliation{$^5$School of Engineering and Information Technology, University of New South Wales Canberra, ACT, 2600, Australia}

%\date{\today}% It is always \today, today,
             %  but any date may be explicitly specified

\begin{abstract}
A general strategy for the realization of electric and magnetic quasi-trapped modes located at the same spectral position is presented. This strategy's application makes it possible to design metasurfaces allowing switching between the electric and magnetic quasi-trapped modes by changing the polarization of the incident light wave. The developed strategy is based on two stages: the application of the dipole approximation for determining  the conditions required for the implementation of trapped modes and the creation of the energy channels for their excitation by introducing a  weak bianisotropy in nanoparticles.  
Since excitation of trapped modes results in a concentration of electric and magnetic energies in the metasurface plane, the polarization switching provides possibilities to change and control the localization and distribution of optical energy at the sub-wavelength scale. We demonstrate a  practical method for spectral tuning of quasi-trapped modes in metasurfaces composed of nanoparticles with a pre-selected shape. As an example, the optical properties of a metasurface composed of silicon triangular prisms are analyzed and discussed.  
\end{abstract}

%\keywords{Suggested keywords}%Use showkeys class option if keyword
                              %display desired
\maketitle

\section{Introduction}

Manipulations and control of strong light-matter interactions in artificial nanoparticle structures are in the focus of current research activities.  
One approach is based on the resonant single-particle responses, including localized surface plasmon and  Mie resonances of metal \cite{schuller2010plasmonics,giannini2011plasmonic,halas2011plasmons} and  dielectric 
\cite{evlyukhin2012demonstration,albella2014electric,yang2018anapole} nanoparticles, correspondingly. Another approach is based on the collective resonances of nanoparticle structures, and their periodic arrays \cite{auguie2008collective,giannini2010lighting,zhang2014strong}. Since resonant responses are associated with the excitation of strong electromagnetic fields in and around the nanoparticles and nanoparticle structures, modern nanophotonics is mainly focused on investigations of resonant light-matter interactions. The development and applications of resonant responses significantly depend on the absorption properties of nanostructures. In contrast to plasmonic structures, where light absorption is one of the main mechanisms of energy losses, resonant dielectric nanostructures allow accumulation and concentration of optical energy at the nanoscale with negligible losses.  The efficiency of light energy concentration in dielectric nanostructures depends on the individual nanoresonators' quality factor and the nanostructure's collective resonances. Recently it has been shown that all-dielectric structures and 2D nanoparticles arrays (metasurfaces) can support ultra-narrow resonances associated with, so-called, trapped modes or bound states in the continuum (BICs) \cite{zhang2013near,Soljacic_NatRevMat_2016,tuz2018high,Koshelev_PhysRevLett_2018,han2019all,koshelev2019meta}. 
The trapped modes or BICs are protected eigenmodes of optical systems remaining perfectly localized without radiation into free space \cite{monticone2019can}. This provides infinite Q-factor resonances and perfect confinement of optical energy and makes these modes very attractive from fundamental and application perspectives \cite{abujetas2019spectral,Li_PhysRevA_2019,han2021extended}.  True optical trapped modes are theoretical objects that can only be realized in ideal lossless infinite structures. In real practical cases, similar states can be obtained by distortion or perturbation of ideal structures' configuration properties \cite{Koshelev_PhysRevLett_2018,fedotov2007sharp}.  In this case, the trapped modes are converted into quasi-trapped modes (quasi-BICs). 
 The excitation of quasi-trapped modes leads to
a high Q-factor resonance and strong near-field enhancement, which may initiate the nonlinear
\cite{Miroshnichenko_AdvSci_2019} and thermal \cite{Miroshnichenko_Small_2019} processes in the nanostructure.

In this paper, a general strategy for realising electric and magnetic quasi-trapped modes located at the same spectral position is developed. We focus on the design and theoretical investigations of metasurfaces allowing the switching between the electric and magnetic quasi-trapped modes by changing the incident light wave's polarization. This opens up new possibilities for the concentration and control of electromagnetic energy at the sub-wavelength scale. We consider quasi-trapped modes, which can also exist in finite-size nanostructures and metasurfaces, which is essential for experimental verification of the obtained theoretical results.

\section{Electric and magnetic trapped modes}

 In order to  realize polarization switching,  the physical mechanisms leading to the existence of electric and magnetic trapped modes should be clarified. Numerical calculations are not suitable for this purpose because of their implicit formalization. Therefore, we apply an analytical approach based on the multipole technique.   In this case,  each particle in the metasurface is replaced by its multipole moments located at the center of mass.  When the particle size is small enough compared to the incident electromagnetic wave's wavelength, the multipole representation can be limited only by the dipole terms.
 The optical properties of such particles are determined by their electric $\hat\alpha^e$ and magnetic $\hat\alpha^m$ dipole polarizability tensors. 
 
 \subsection{Trapped modes as symmetry protected BICs}
 We consider an infinite metasurface with a square elementary cell composed of identical dipole particles irradiated by a monochromatic light plane wave at normal incidence. In this conditions, all particles have the same electric $\bf p$ and magnetic $\bf m$ dipole moments satisfying  the following equations \cite{evlyukhin2010optical}
\begin{eqnarray}
    (\hat {\bf 1}-\hat\alpha^e\hat S^e){\bf p}=\varepsilon_0\varepsilon_S\hat\alpha^e{\bf E}\:,\label{e1}\\
    (\hat {\bf 1}-\hat\alpha^m\hat S^m){\bf m}=\hat\alpha^m{\bf H}\:,\label{e2}
\end{eqnarray}
where $\bf E$ and $\bf H$ are the external electric and magnetic fields of the incident wave in the metasurface plane, $\varepsilon_0$ and $\varepsilon_S$ are the vacuum permittivity and relative dielectric permittivity of the surrounding medium, $\hat S^e$ and $\hat S^m$ are the tensors of electric and magnetic dipole sums accounting for the electromagnetic interaction between the dipoles arranged into a periodic lattice of the metasurface, $\hat {\bf 1}$ is the unit $3\times3$ tensor. 
Note that the above equations correspond to a local response theory where the dipole moments are determined by local fields and do not depend on their spatial derivatives \cite{bobylev2020nonlocal}  (an expended model with nonlocal bianisotropic responses will be given below).     
In the international system of units, one has $\hat S^m=\hat S^e\equiv\hat S$. These dipole sums depend on the metasurface period $d$, the incident wavelength $\lambda$ in the surrounding medium,  but are independent on particle characteristics and properties, so that  $\hat S=\hat S(d,\lambda)$.  
Assuming that the metasurface lattice is located in the $xy$ plane at  $z = 0$ of the Cartesian coordinate system, due to the lattice
periodicity in the $x$ and $y$ directions, the tensor $\hat S$ becomes diagonal  with the non-zero elements $S_{xx}$, $S_{yy}$, and $S_{zz}$ \cite{evlyukhin2010optical,babicheva2019analytical}. Moreover, for metasurfaces with a square elementary cell the condition  $S_{xx}=S_{yy}\equiv S_{\parallel}$ is satisfied. 
We assume also that the particle polarizabilities are diagonal matrices with the elements $\alpha_{xx}^e$, $\alpha_{yy}^e$, and $\alpha_{zz}^e$ for $\hat \alpha^e$ and $\alpha_{xx}^m$, $\alpha_{yy}^m$, and $\alpha_{zz}^m$ for $\hat \alpha^m$. In general, these elements' values can depend on the particle shape, size, and dielectric permittivity, and the incident light wavelength.

The trapped modes (symmetry-protected BICs) of the metasurfaces correspond to solutions of Eqs. (\ref{e1}) and (\ref{e2}) in the absence of external fields \cite{abujetas2020coupled}. These modes are  localized
within the metasurface and are not able to emit energy in the far
field zone \cite{hsu2016bound}.  Assuming zero external fields,  we obtain from (\ref{e1}) and (\ref{e2})
\begin{eqnarray}
   \left(\frac{1}{\alpha_{\parallel}^e} -S_{\parallel}\right) p_{\parallel}=0\:,\quad
   \left(\frac{1}{\alpha_{\parallel}^m} -S_{\parallel}\right) m_{\parallel}=0\:,\label{e3}\\
    \left(\frac{1}{\alpha_{zz}^e} -S_z\right) p_z=0\:,\quad
   \left(\frac{1}{\alpha_{zz}^m} -S_z\right) m_z=0\:,\label{e4}
\end{eqnarray}
where $\alpha_{\parallel}$, $p_{\parallel}$, and $m_{\parallel}$ are the corresponding in-plane (in the $xy$-plane) components of the polarizabilities and dipole moments.  As has been discussed in Subsection VIII.B of  Ref. \cite{babicheva2021multipole}, in a non-diffractive regime, when the metasurface period is smaller than the incident wavelength, Eqs. (\ref{e3}) have only trivial solutions ($p_{\parallel}=0$ and $m_{\parallel}=0$).  Therefore,  no trapped modes can be associated solely with the in-plane components of the dipole moments. Unlike  Eqs. (\ref{e3}),  solutions to Eqs. (\ref{e4}) may differ from zero in the case of non-absorbing dipole scatterers \cite{babicheva2021multipole}. Conditions for the existence of non-trivial solutions can be obtained from the two independent equations 
\begin{eqnarray}
\frac{1}{\alpha_{zz}^e} -S_z=0\: \Rightarrow \frac{1}{\alpha_{zz}^e} =S_z\:,\label{e5}\\
\frac{1}{\alpha_{zz}^m} -S_z=0\:\Rightarrow \frac{1}{\alpha_{zz}^m} =S_z\:.\label{e6}
\end{eqnarray}
Equation (\ref{e5}), involving only electric dipole  coupling, corresponds to an {\it electric trapped mode}, and Eq. (\ref{e6}), involving only magnetic dipole coupling,  corresponds to a {\it magnetic trapped mode}. Spectral positions of the electric and magnetic trapped modes can be found using Eqs. (\ref{e5}) and  (\ref{e6}), respectively. From these equations, one can get an important conclusion:  to have the trapped modes  at the same spectral position, the following condition  
\begin{equation}\label{e7}
   \frac{1}{\alpha_{zz}^e}= \frac{1}{\alpha_{zz}^m}=S_z
\end{equation}
should be satisfied.

\begin{figure}[htbp]
\centering
%\fbox{\includegraphics[width=\linewidth]{D_6.eps}}
\includegraphics[width=0.9\linewidth]{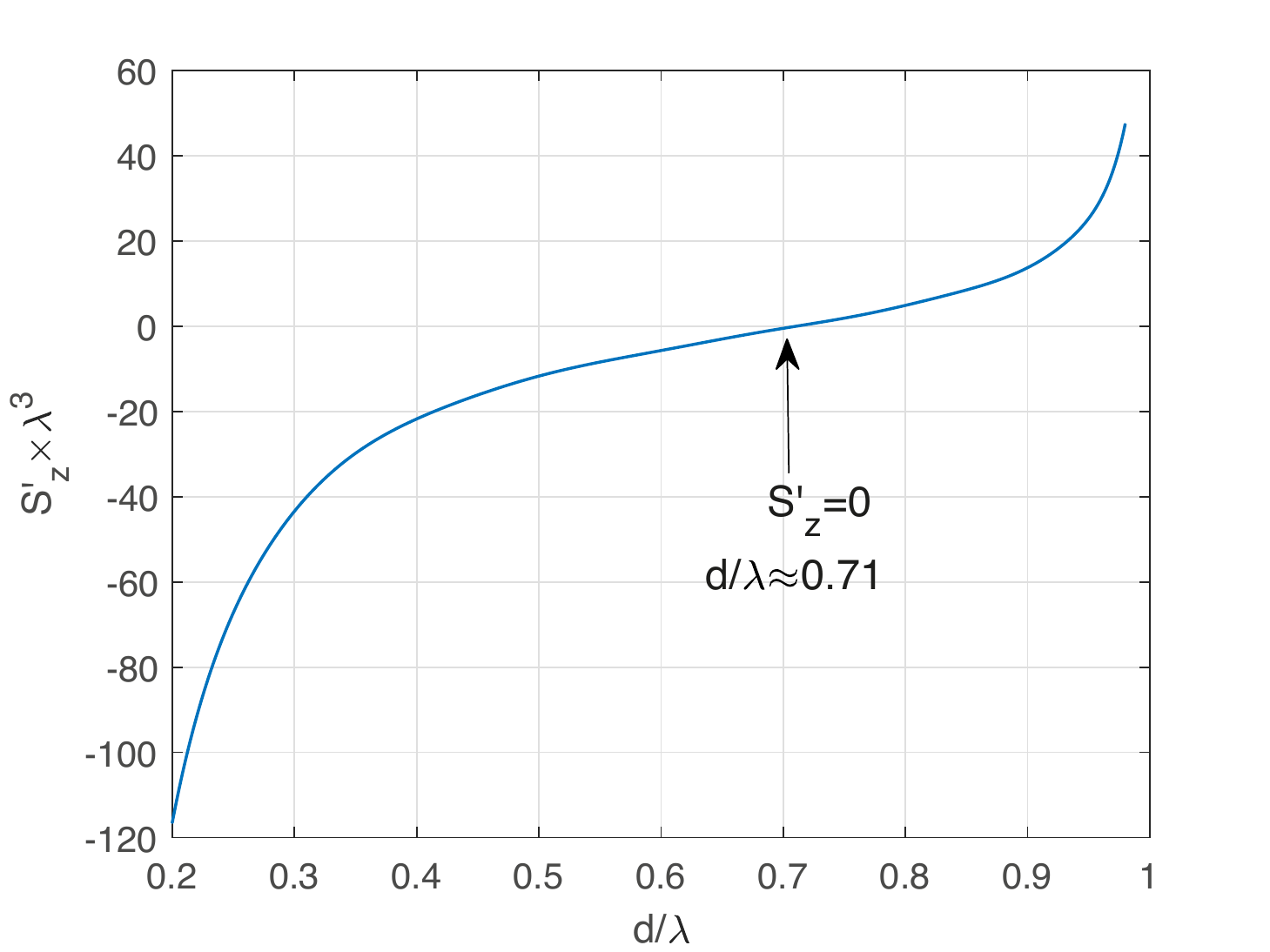}
\caption{Real part $S_z'$ of the dipole sum $S_z$ calculated for different periods $d$ of metasurfaces with square ($d\times d$) elementary cells and for a fixed wavelength ($\lambda>d$) corresponding to the wave number $k=2\pi/\lambda$ in the medium surrounding metasurfaces.}
\label{fig1}
\end{figure}

Expressions for the imaginary $S_z^{''}$ and real $S_z^{'}$ parts of $S_z=S_z^{'}+iS_z^{''}$ are presented by 
\begin{equation}
\label{eq:S1}
S'_z=\frac{k^2}{4\pi}\sum_{i=0,\pm1}^{\pm\infty}\sum_{j=0,\pm1}^{\pm\infty}\left(\frac{\cos kd_{ij}}{d_{ij}}-\frac{\sin kd_{ij}}{kd_{ij}^2}-\frac{\cos kd_{ij}}{k^2d_{ij}^3}\right),
\end{equation}
and
\begin{equation}
%\begin{split}
S''_z=\frac{k^2}{4\pi}\sum_{i=0,\pm1}^{\pm\infty}\sum_{j=0,\pm1}^{\pm\infty}\left(\frac{\sin kd_{ij}}{d_{ij}}+\frac{\cos kd_{ij}}{kd_{ij}^2}-\frac{\sin kd_{ij}}{k^2d_{ij}^3}\right)\:,
%\end{split}
\label{eq:S11}
\end{equation} 
where $d_{ij}=d\sqrt{i^2+j^2}$,   $k$ is the wave number in the surrounding medium, and the term with $d_{00}=0$ is  excluded from the sums. Note that Eqs. (\ref{eq:S1}) and (\ref{eq:S11}) contain explicit sum limits compared to similar equations in Ref. \cite{evlyukhin2020bianisotropy}. Figure \ref{fig1} demonstrates the period behavior of normalized $S_z^{'}$ calculated  using a method described in \cite{babicheva2021multipole}. Note that the presentation in Fig. \ref{fig1} does not change when the refractive index of the surrounding homogeneous medium changes.    By direct analytical calculations, it is possible to show that in non-diffractive case (when $\lambda>d$ or $kd<2\pi$) the imaginary part has the following form
\begin{equation}
    S_z^{''}=-k^3/(6\pi).
\end{equation}
On the other hand, for a dielectric non-absorbing particle having diagonal tensors of the electric $\hat\alpha^e$ and magnetic $\hat\alpha^m$ dipole polarizabilties, we have from the optical theorem  \cite{babicheva2021multipole,alu2006theory}
\begin{equation}\label{e8}
   \Im\frac{1}{\alpha_{zz}^e}=\Im \frac{1}{\alpha_{zz}^m}=-k^3/(6\pi)\:.
\end{equation}
Thus, in a non-diffractive case, the condition (\ref{e7}) for the imaginary parts are satisfied in the spectral range where $kd<2\pi$. In order to fulfil the condition (\ref{e7})
 for the real parts, we can apply different tuning procedures.  We can suggest three possibilities.
 \\
 i) Predetermined particle. One can find the spectral point ($\lambda'$), where $\Re(\lambda'^3/\alpha_{zz}^e)=\Re(\lambda'^3/\alpha_{zz}^m)\equiv A$ for a given single particle, and then, using the curve in Fig. \ref{fig1}, one finds the value $S_z'\times\lambda'^3=A$ and determines the corresponding period $d/\lambda'$ of the metasurfaces composed of such particles and supporting the electric and magnetic trapped modes at the same wavelength.\\
 ii) Predetermined wavelength. For a given wavelength $\lambda'$, one chooses the geometry and size of dielectric particles for which  $\Re(\lambda'^3/\alpha_{zz}^e)=\Re(\lambda'^3/\alpha_{zz}^m)\equiv A$ is fulfilled, and then, using the curve in Fig. \ref{fig1}, one finds the value $S_z'\times\lambda'^3=A$ and determines the corresponding period $d/\lambda'$ of the metasurfaces.\\ 
 iii) Particle tuning  (an especially resonant case). It has been shown \cite{evlyukhin2011multipole,staude2013tailoring} that by tuning the geometrical parameters of single particles, it is possible to get excitation of their electric and magnetic dipole resonances at a certain spectral position with $\lambda_0$, where   $\Re(1/\alpha_{zz}^e)=\Re(1/\alpha_{zz}^m)= 0$. Therefore, for the realization of both trapped modes at this resonant condition, we can only consider the spectral position in Fig. \ref{fig1} corresponding to $S_z'=0$. As a result, we get the period  $d\approx0.71\lambda_0$ for the metasurface composed of such particles and supporting the electric and magnetic trapped modes at $\lambda_0$. It is important to note that the "particle tuning" case can also be used to determine the periods of metasurfaces supporting a trapped mode of the only electric or magnetic type. To accomplish this, one should consider the electric or magnetic dipole resonant wavelength $\lambda_0^R$ separately so that the corresponding period $d$ will be equal to $0.71\lambda_0^R$ (see Fig. \ref{fig1}).

It may happen that the period value $d$, determined by the above procedures,   is smaller than the lateral (in-plane) dimensions of particles, and hence such metasurface cannot exist. In this case, the particles with other material parameters (with a higher value of the refractive index) should be considered. If the non-local contributions to the dipole moments are not negligible, Eqs. (\ref {e1}) and (\ref {e2}) will change (see the next subsection)  due to the additional terms with the space derivatives of electromagnetic fields, and consequently, conditions for the realization of trapped modes may be modified. However, if the dipole approximation is applicable, the non-local corrections of the dipole $zz$-polarizabilities  are still small in the optical range  \cite{bobylev2020nonlocal}.
 
\subsection{Quasi-trapped modes}

The considered trapped modes correspond to the symmetry protected BICs \cite{Li_PhysRevA_2019,abujetas2020coupled} that cannot radiate in the far-field zone and, therefore, cannot be excited by external incident plane waves. To excite them from the far field,   one has to perturb the symmetry properties of the system. One of the frequently used ways is introducing a regular asymmetric parameter characterizing the deviation of the particle shape from its symmetrical counterpart \cite{tuz2018high}. As a result of such transformation of particles,  their dipole polarizability tensors get bianisotropic off-diagonal elements \cite{evlyukhin2020bianisotropy} as a consequence of the non-locality of their optical response. In this case, Eqs. (\ref{e1}) and (\ref{e2}) are replaced by
\begin{eqnarray}
    (\hat {\bf 1}-\hat\alpha^e\hat S){\bf p}-c_S^{-1}\hat\alpha^{em}\hat S{\bf m}&=&\varepsilon_0\varepsilon_S\hat\alpha^e{\bf E}+c_S^{-1}\hat\alpha^{em}{\bf H}\:,\label{e9}\nonumber\\
    \\
    -c_S\hat\alpha^{me}\hat S{\bf p}+(\hat {\bf 1}-\hat\alpha^m\hat S){\bf m}&=&c_S\hat\alpha^{me}{\bf E}+\hat\alpha^m{\bf H}\:,\label{e10}
\end{eqnarray}
where $c_S$ is the light velocity (in the surrounding medium with $\varepsilon_S$), $\hat\alpha^{em}$ and $\hat\alpha^{me}$ are the tensors of bianisotropic polarizabilities which allow excitation of the particle electric dipole by the magnetic field and the particle magnetic dipole by the electric field, correspondingly. Remind that Eqs. (\ref{e9}) and (\ref{e10}) are written for normal incidence of the external plane waves. 

\begin{figure}[htbp]
\centering
%\fbox{\includegraphics[width=\linewidth]{D_6.eps}}
\includegraphics[width=0.7\linewidth]{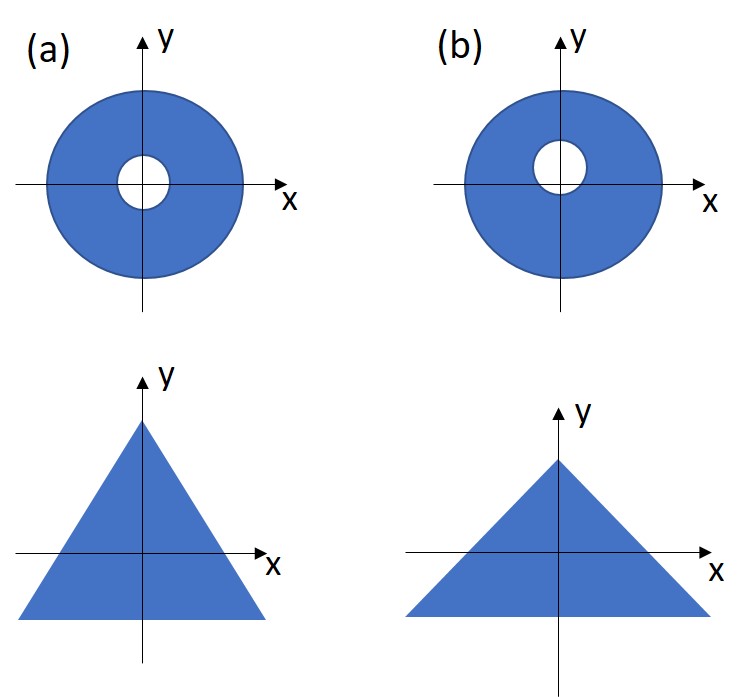}

\includegraphics[width=0.7\linewidth]{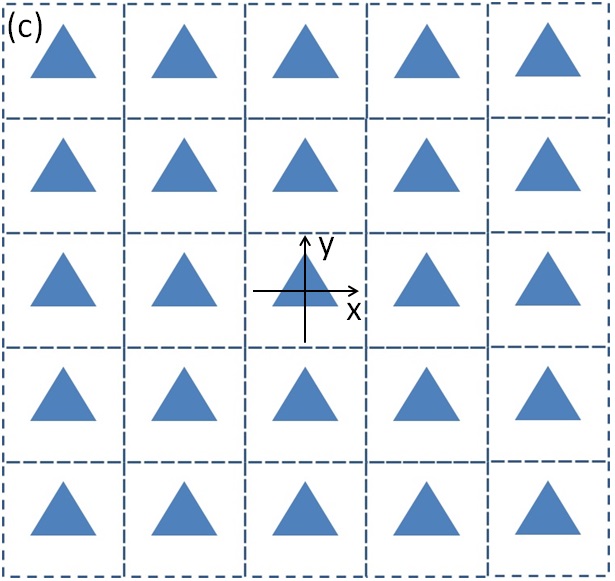}
\caption{Top view of disks with a hole and triangular prisms: (a) with and (b) without rotational symmetry around the $z$-axis. (c) Top view  of a fragment of the metasurface composed of silicon prisms.}
\label{fig2}
\end{figure}

The trapped modes considered above (or the symmetry-protected BICs) arise due to the electromagnetic coupling between the corresponding out-of-plane dipole moments, i.e. oriented perpendicular to the metasurface plane. However, such dipole moments cannot be excited by normally incident plane waves if the metasurface consists of particles with in-plane rotational symmetry, such as spheres, disks, cubes, cones, symmetric prisms, etc. This corresponds to the practical rule: a light plane  wave does not excite the longitudinal (co-linear to wave propagation) component of dipole moments of a  scattering particle if the direction of its propagation coincides with the rotational symmetry axis of the particle. Therefore, in order to excite  the metasurface's trapped modes, the particles' in-plane rotational symmetry must be broken. Two examples of symmetry breaking for disk and prismatic particles are shown in Fig. \ref{fig2}. Introduction of such symmetry breaking leads to the appearance of bianisotropic off-diagonal polarizabilities $\alpha_{zx}^{me}$ and $\alpha_{zx}^{em}$ \cite{evlyukhin2020bianisotropy}. Applying the 
magnetoelectric optical theorem for particles with bianisotropic properties \cite{sersic2011magnetoelectric}, Eq. (\ref{e8}) is replaced with
\begin{eqnarray}
\Im\frac{1}{\alpha_{zz}^m}=-\frac{k^3}{6\pi}\left(1+\frac{|\alpha_{zx}^{me}|^2}{|\alpha_{zz}^{m}|^2}\right)\:, \label{e11}\\
\Im\frac{1}{\alpha_{zz}^e}=-\frac{k^3}{6\pi}\left(1+\frac{|\alpha_{zx}^{em}|^2}{|\alpha_{zz}^{e}|^2}\right)\:. \label{e12}
\end{eqnarray}
This indicates a distortion of the existence conditions for
trapped modes compared to a metasurface with symmetric
particles. As a result, the pure trapped modes are transformed into quasi-trapped modes with a finite spectral width.    Nonetheless, we may expect that, because of a weak asymmetric perturbation, the second terms in the brackets of Eqs. (\ref{e11}) and  (\ref{e12}) remain small, and the spectral positions of the magnetic and electric quasi-trapped modes can be estimated from the resonant conditions 
\begin{equation}\label{eq16}
    \Re\frac{1}{\alpha_{zz}^m}=S'\quad {\rm and}\quad \Re\frac{1}{\alpha_{zz}^e}=S'\:,
    \end{equation}
respectively. Note that the quasi-trapped modes' excitation is also accompanied by strong electric and magnetic near fields in the metasurface plane.

In the following sections, we show how the discussed approach can be applied to develop metasurfaces composed of silicon triangle prisms  (see Fig. \ref{fig2}c) with a possibility of polarization switching between the electric and magnetic quasi-trapped modes in the near-infrared spectral range.  To do this, we will apply the first (i) possibility described above.

\section{Optical response of  silicon nanoprisms}
\label{sec:examples}

We consider silicon nanoparticles in the form of triangular prisms with the relative dielectric permittivity $\varepsilon_p=12.67$. The dimensional parameters are chosen to obtain the prism's electric and magnetic dipole responses in the near-infrared spectral range, where light absorption in silicon is negligible \cite{palik1998handbook}.  In this paper, the total electric field $\bf E$ in the prisms and the corresponding scattering cross sections (SCSs) are calculated
numerically using the finite element method (FEM) in COMSOL Multiphysics. Then the induced polarization ${\bf P}=\varepsilon_0(\varepsilon_p-\varepsilon_S) {\bf E}$ or the displacement current density ${\bf j}=-i\omega {\bf P}$ can
be calculated using the obtained field, where $\omega$ is the angular frequency and the time dependence $\exp(-i\omega t)$ is considered.  After that, the multipole moments and their contributions to the SCS are calculated following the method described in Ref. \cite{terekhov2017multipolar}.  The Cartesian multipole moments of the scatterers are calculated following their definitions presented in Ref. \cite{alaee2018electromagnetic,evlyukhin2019multipole}.   The dipole polarizabilities are calculated using the approach described in Ref. \cite{evlyukhin2020bianisotropy}. The dielectric constant $\varepsilon_S=1$, and the electric field amplitude of the incident linear-polarized plane wave is chose to be equal  1 $V/m$ for all simulations.

\begin{figure}[htbp]
\centering
%\fbox{\includegraphics[width=\linewidth]{D_6.eps}}
\includegraphics[width=1\linewidth]{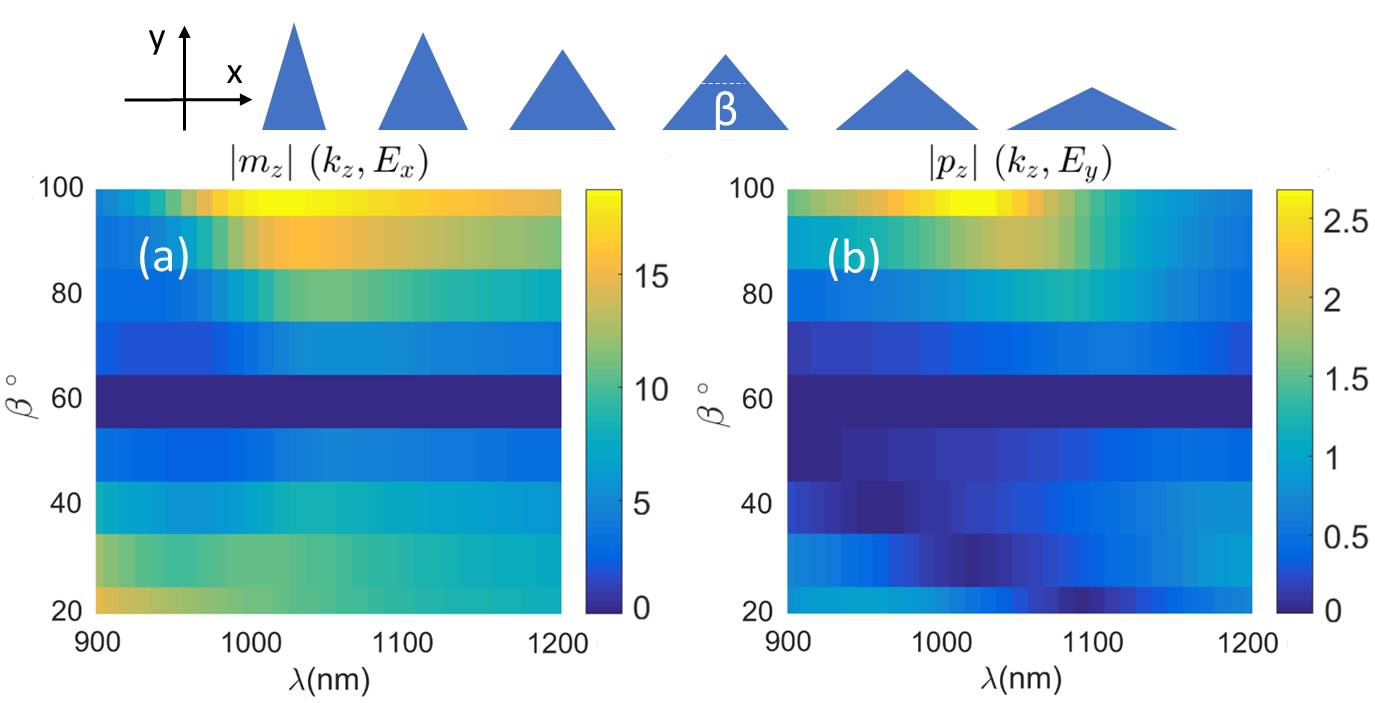}
\caption{Absolute values (in arbitrary units) of the longitudinal (co-linear to wave propagation) $z$-component of the (a) magnetic $m_z$ and (b) electric $p_z$ dipole moments for silicon prisms irradiated by plane waves propagating along the $z$-axis and having electric polarization along (a) $x$-axis and (b) $y$-axis. For clarity, the inset above the main pictures illustrates the Cartesian coordinate system's orientation and changes of the prism bases with the increasing angle $\beta$.  The dipole moments are calculated with respect to the prism center of mass. }
\label{fig3}
\end{figure}

Figure \ref{fig3} presents spectra of the longitudinal $z$-components of the magnetic and electric dipole moments excited in the prisms with different base angles $\beta$ (see the inset in Fig. \ref{fig3})  by plane waves propagating along the $z$-axis  normally to the prism base. All prisms have the same height and volume (the corresponding  parameters for $\beta=40^{\rm o}$ are shown below).   One can see that the dipole $z$-component  can be excited in all prisms except that with the equilateral triangle base ($\beta=60^{\rm o}$). Notably, the type (electric or magnetic) of the dipole $z$-component is determined by the incident wave's polarization. For the electric- (magnetic-) polarization directed perpendicular to the bisector of the $\beta$ angle, only the longitudinal $z$-component of the magnetic (electric) dipole moment is excited.

In the following simulations we assume that the prism bases are isosceles triangles with  $\beta=40^{\rm o}$ providing  conditions for excitation of the  $z$-component of electric or magnetic dipole moments by  light plane waves propagating along the $z$-axis.
In order to have the electric and magnetic dipole resonances in the spectral range $\lambda=900 -1200$ nm, we choose silicon prisms with the following geometric parameters: the prism height $H=$300 nm, the radius of the circumscribed circle of the base triangle $R=$222 nm, and the angle $\beta=40^{\rm o}$. 

\begin{figure}[htbp]
\centering
\includegraphics[width=0.45\linewidth]{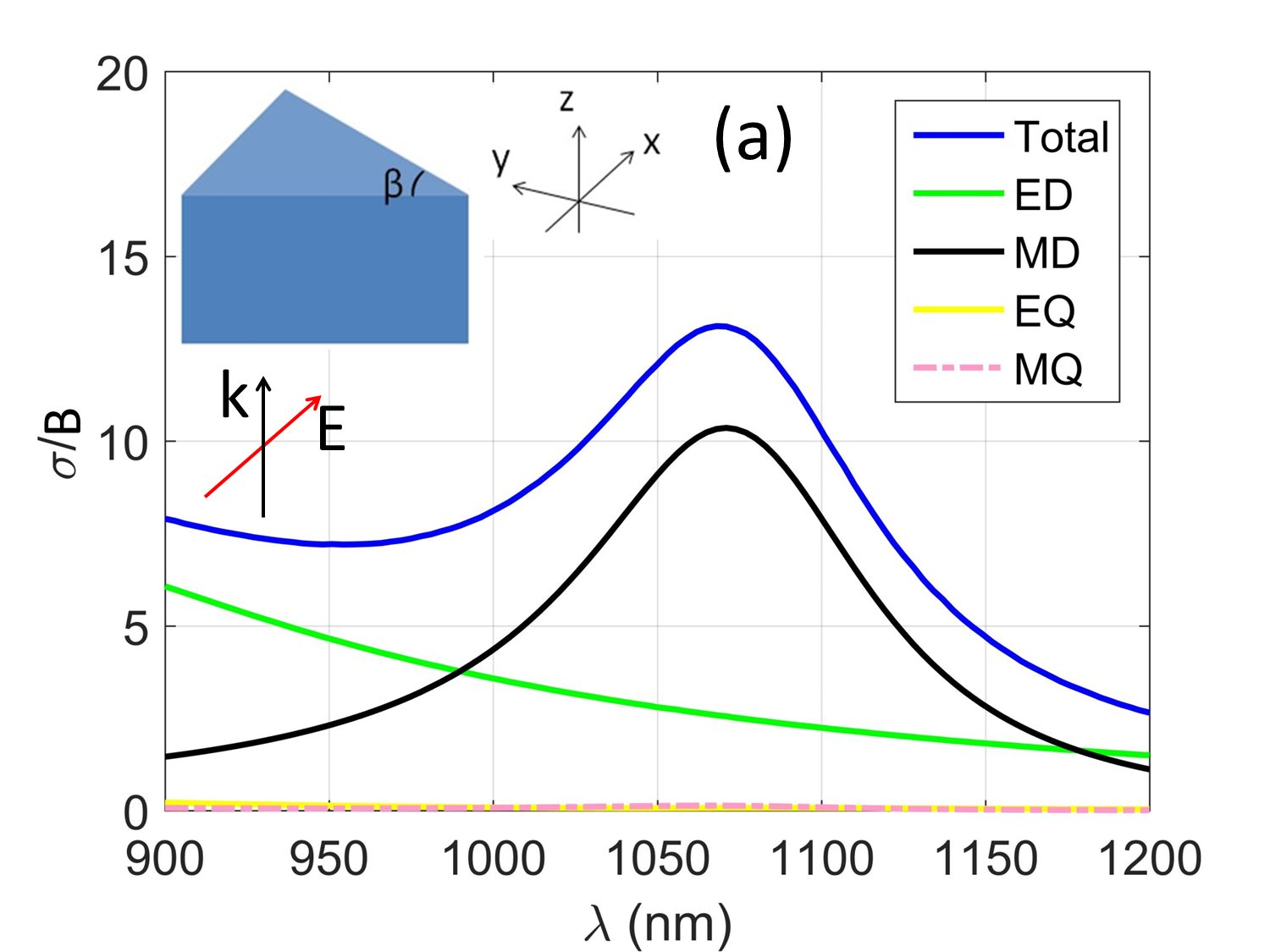}
\includegraphics[width=0.45\linewidth]{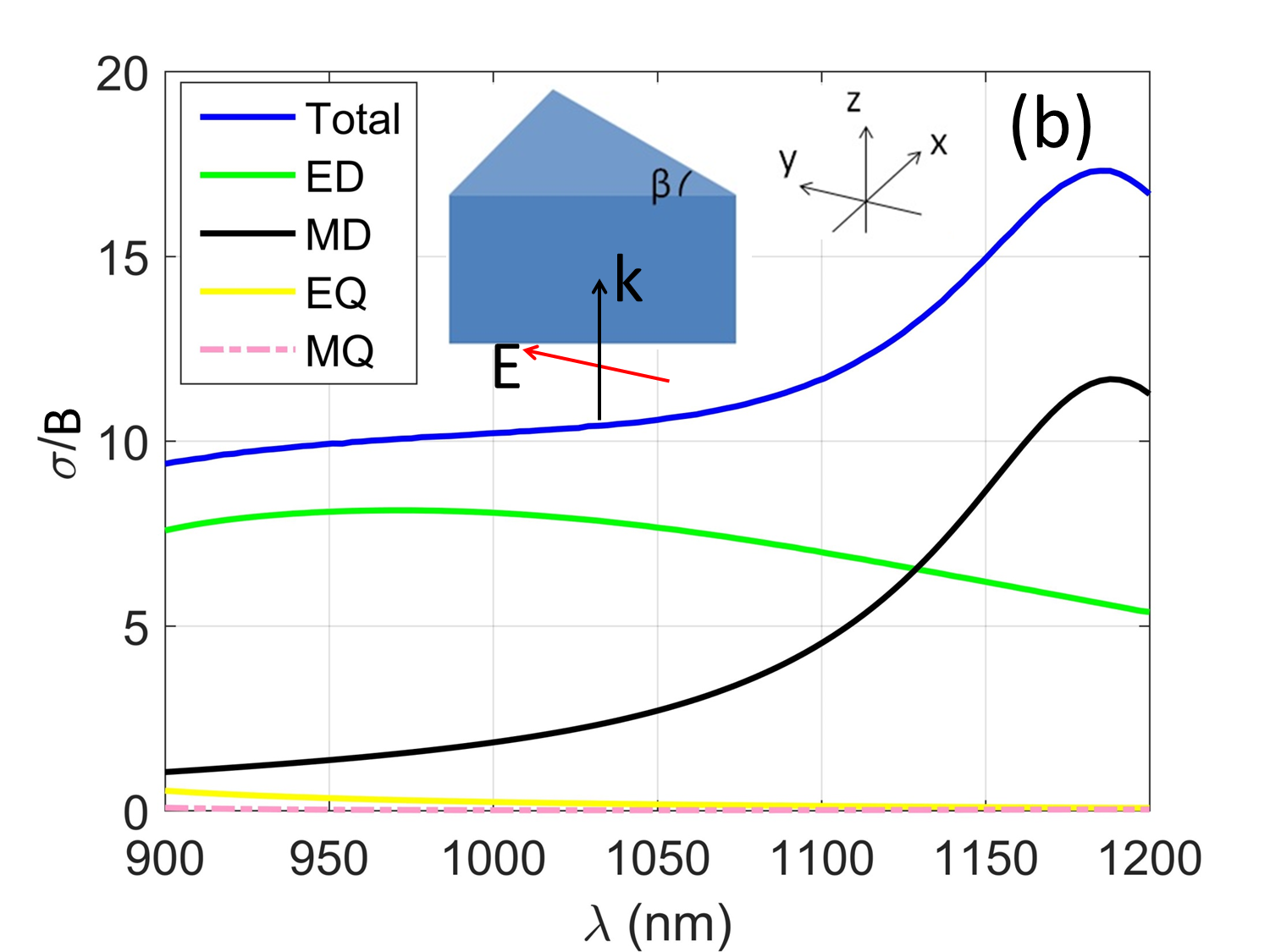}

\includegraphics[width=0.45\linewidth]{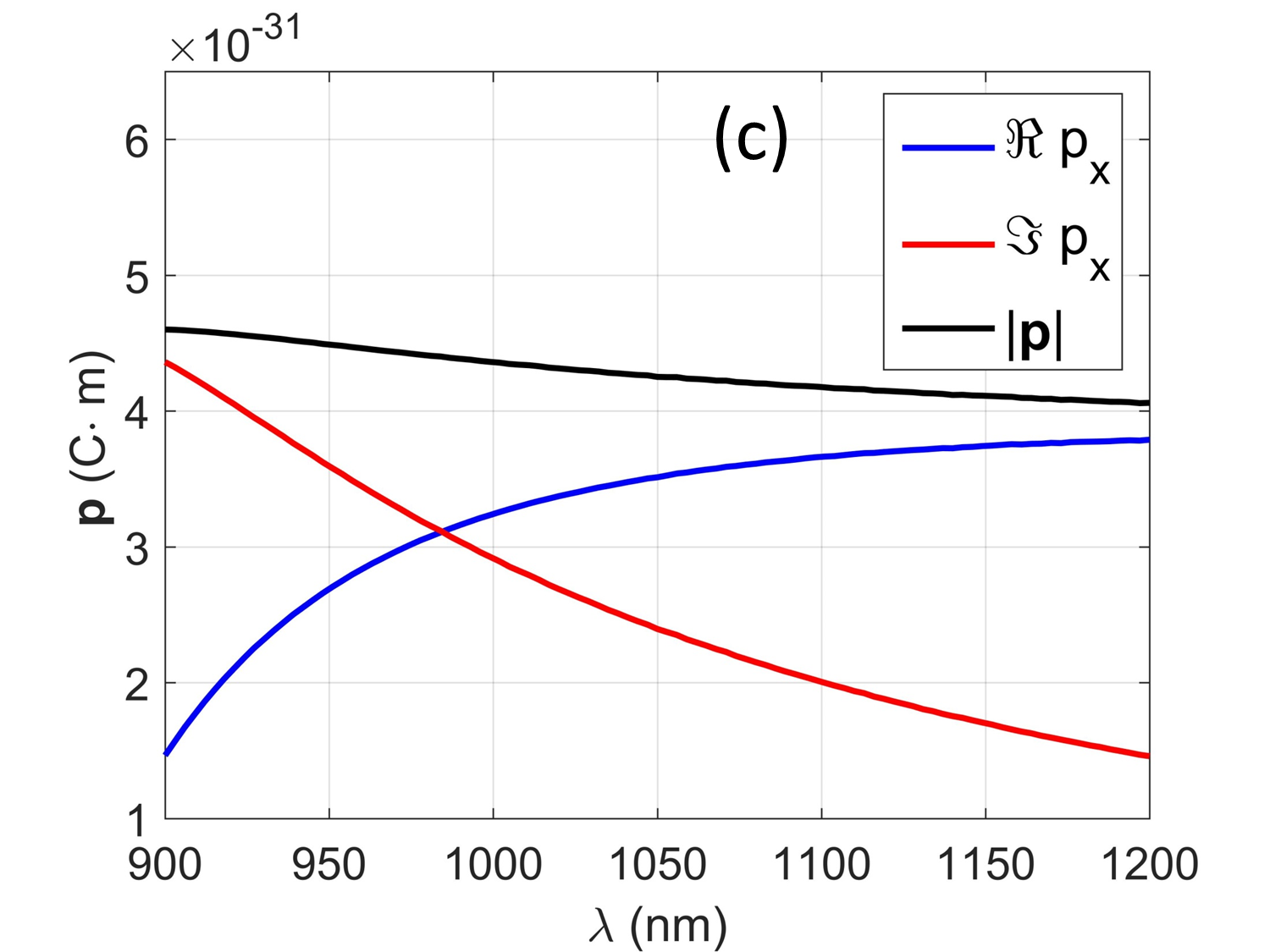}
\includegraphics[width=0.45\linewidth]{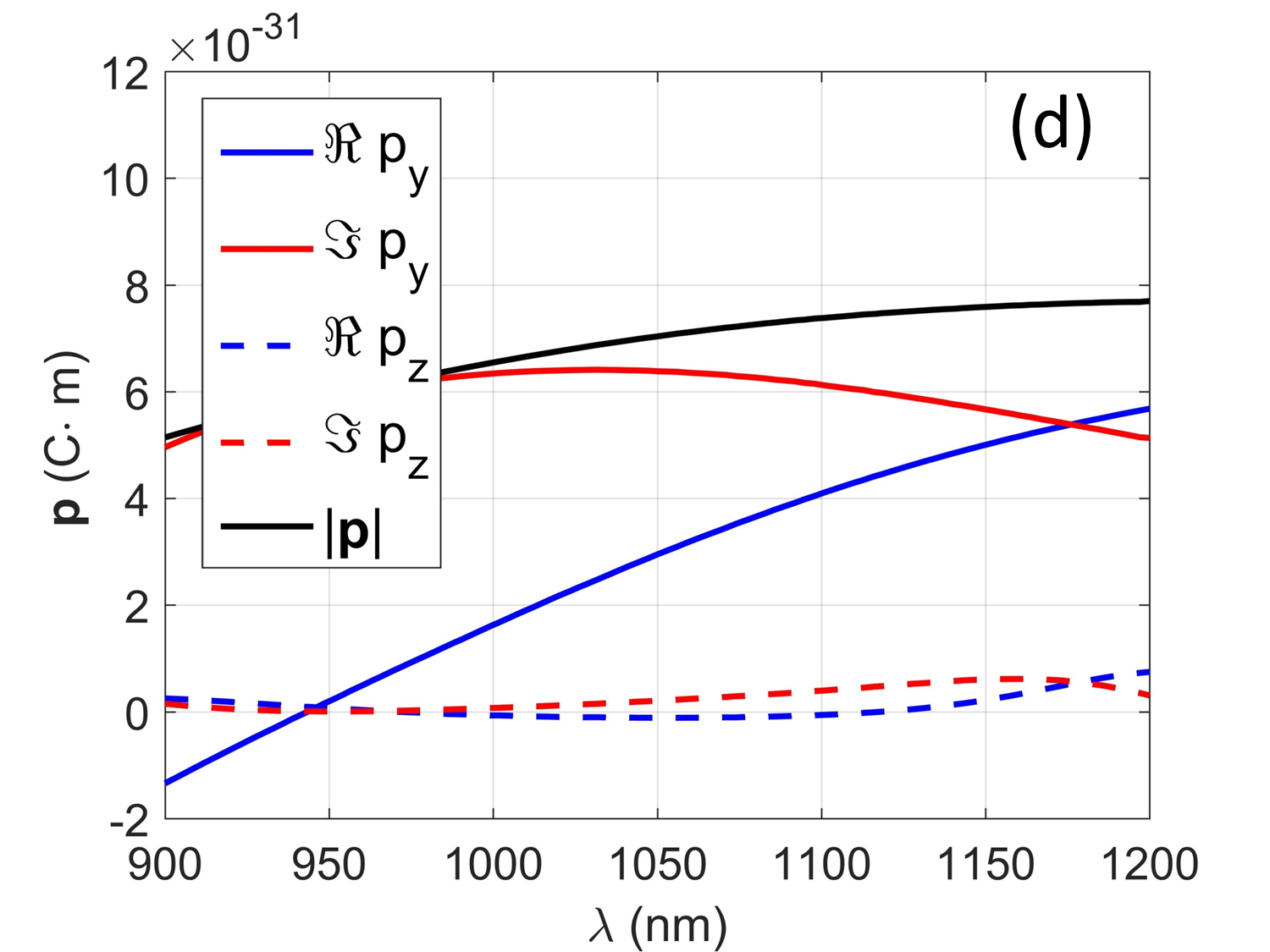}

\includegraphics[width=0.45\linewidth]{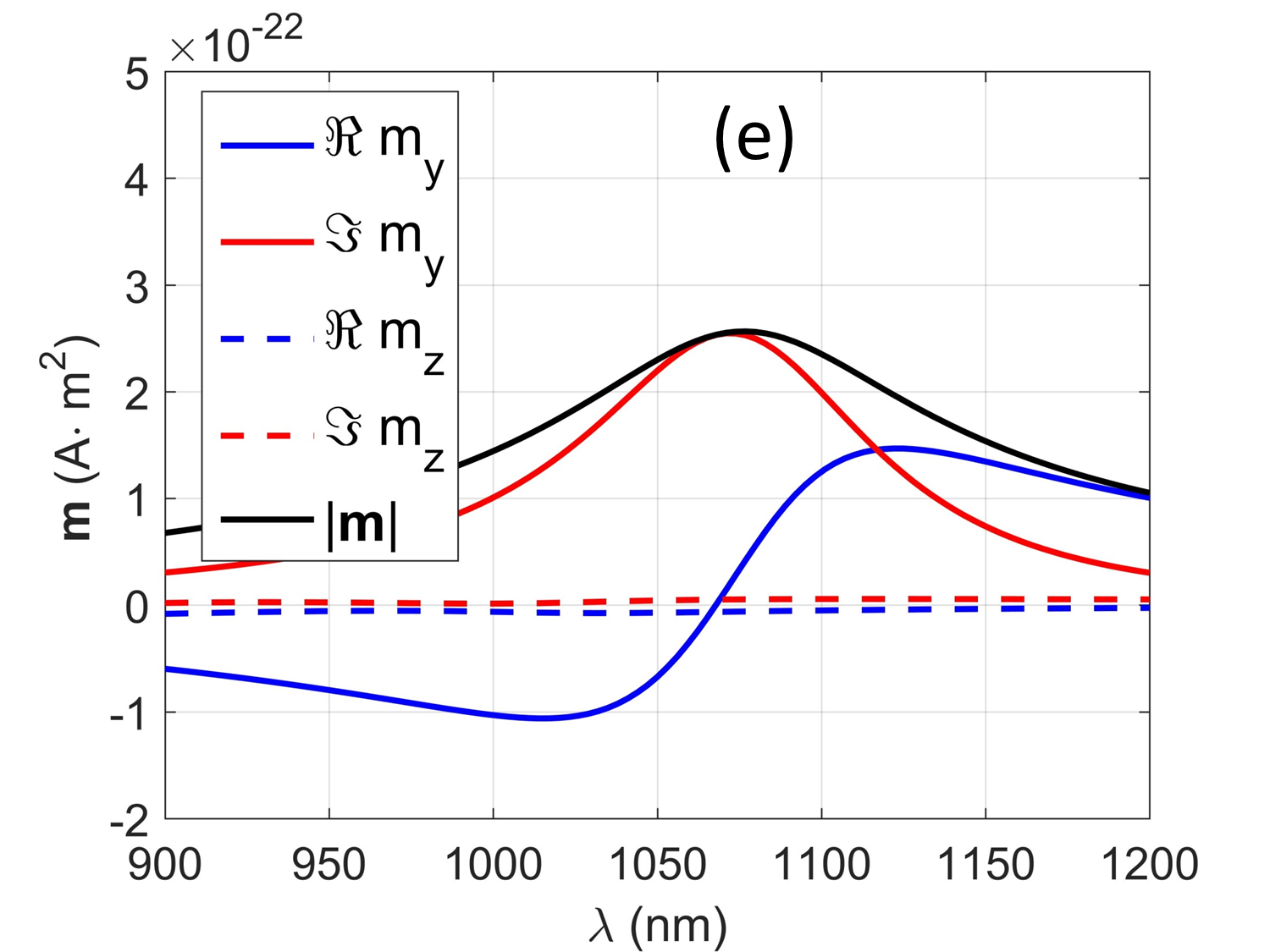}
\includegraphics[width=0.45\linewidth]{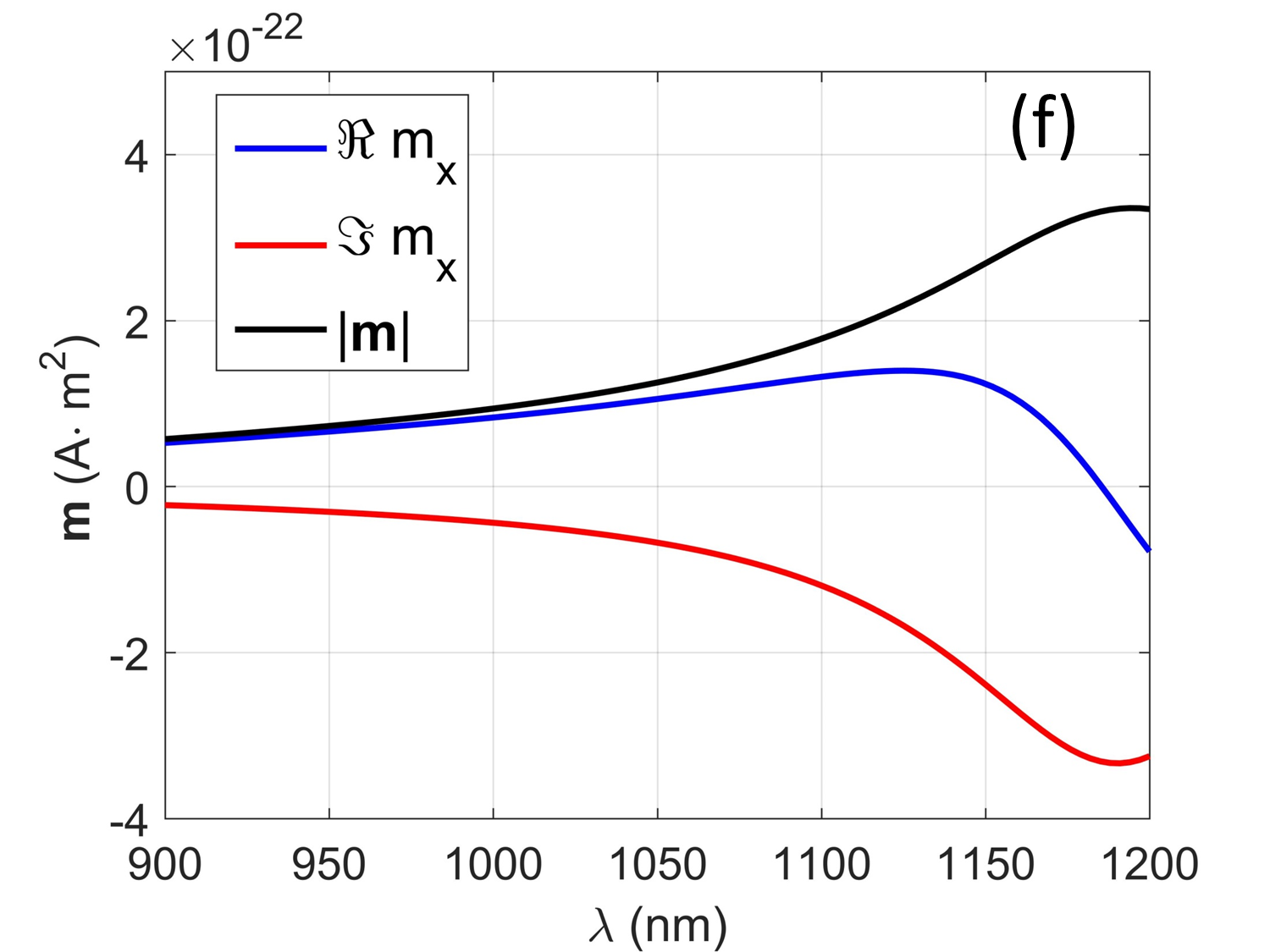}
\caption{(a) and (b) Spectra of the normalized scattering cross sections $\sigma/B$ (where $B$ is the area of the prism base) with corresponding multipole decomposition calculated for the silicon prism. 
Irradiation conditions are shown in the insets. Components of the electric (c) and magnetic (e)  dipole moments excited in the prism at the irradiation condition shown in to the inset in (a).  Components of the electric (d) and magnetic (f) dipole moments excited in the prism at the irradiation condition sown in the inset in (b).}
\label{fig4_1}
\end{figure}

In Fig. \ref{fig4_1}a and Fig. \ref{fig4_1}b one can see that the SCSs are determined by  only contributions of the electric ED and magnetic MD dipole terms, the contributions of the electric EQ and magnetic MQ  quadruple terms are negligible. In accordance with the results in Fig. \ref{fig3} the waves polarized along the $x$-axis (perpendicular to the bisector of the $\beta$ angle) generate the ED moment $p_x$ only along the $x$-axis ( see Fig. \ref{fig4_1}c), whereas the MD moment has two components: $m_y$ along the incident magnetic field and  $m_z$ along the wavevector of the incident light wave (see Fig. \ref{fig4_1}e). 
For the electric $y$-polarization we have the MD moment $m_x$ only along the $x$-axis (Fig. \ref{fig4_1}f) and the ED moment has two components $p_y$  and  $p_z$  (see Fig. \ref{fig4_1}d). The longitudinal dipole  components $m_z$ and $p_z$ are excited due to the bianisotropic  response of the prism with polarizabilties $\alpha_{zx}^{me}$ and $\alpha_{zx}^{em}$. 

\begin{figure}[htbp]
\centering
\includegraphics[width=0.45\linewidth]{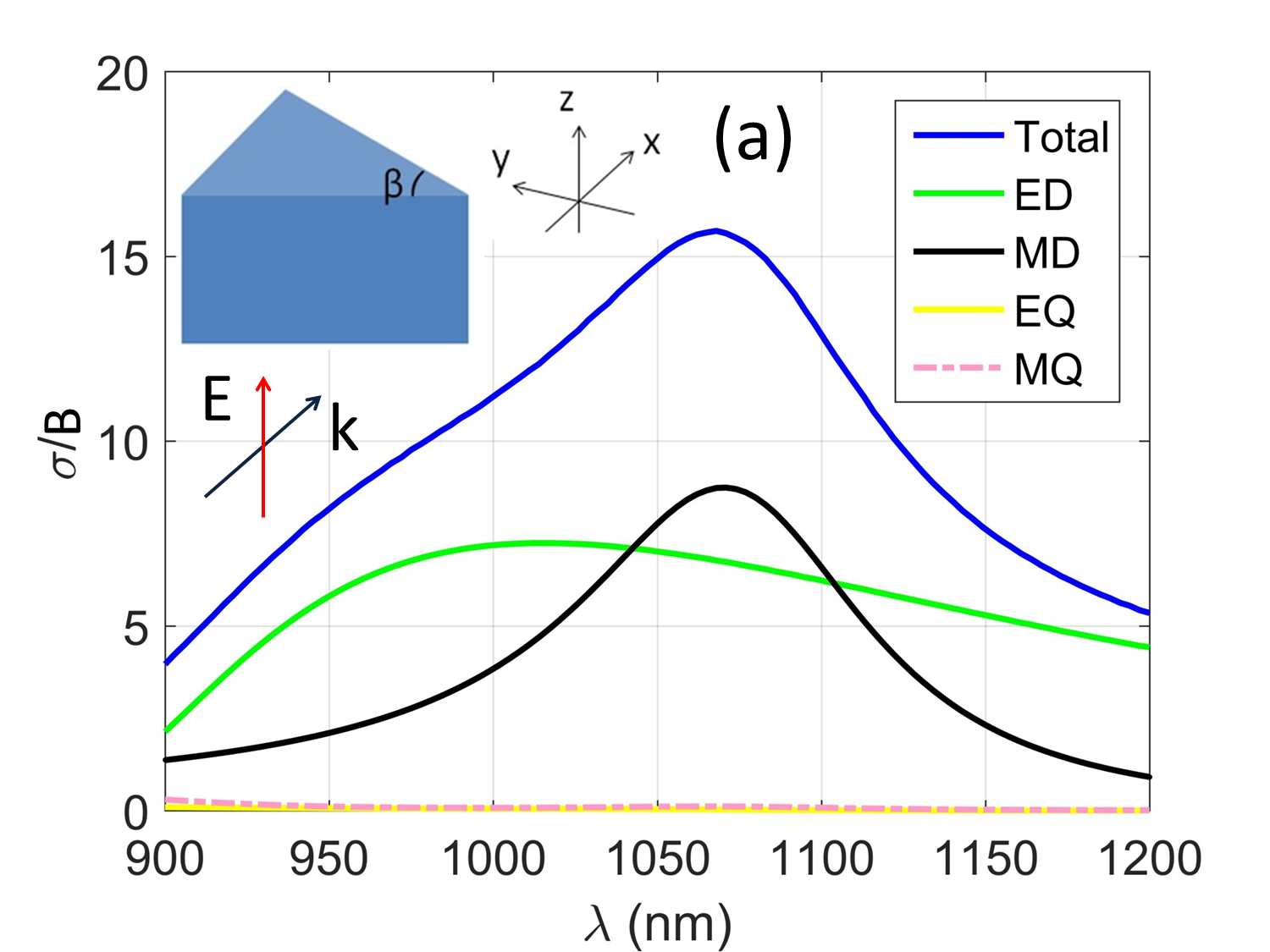}
\includegraphics[width=0.45\linewidth]{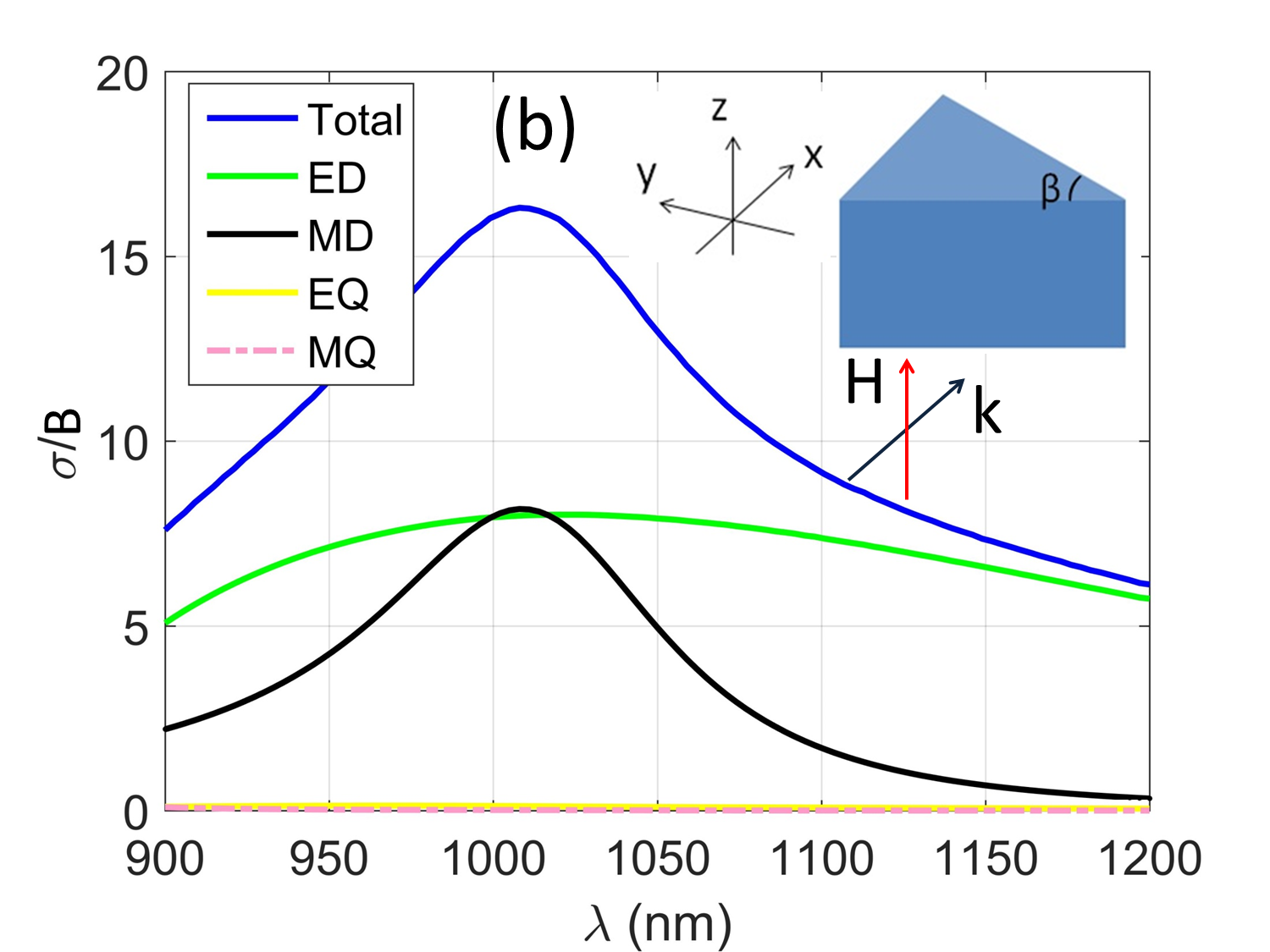}

\includegraphics[width=0.45\linewidth]{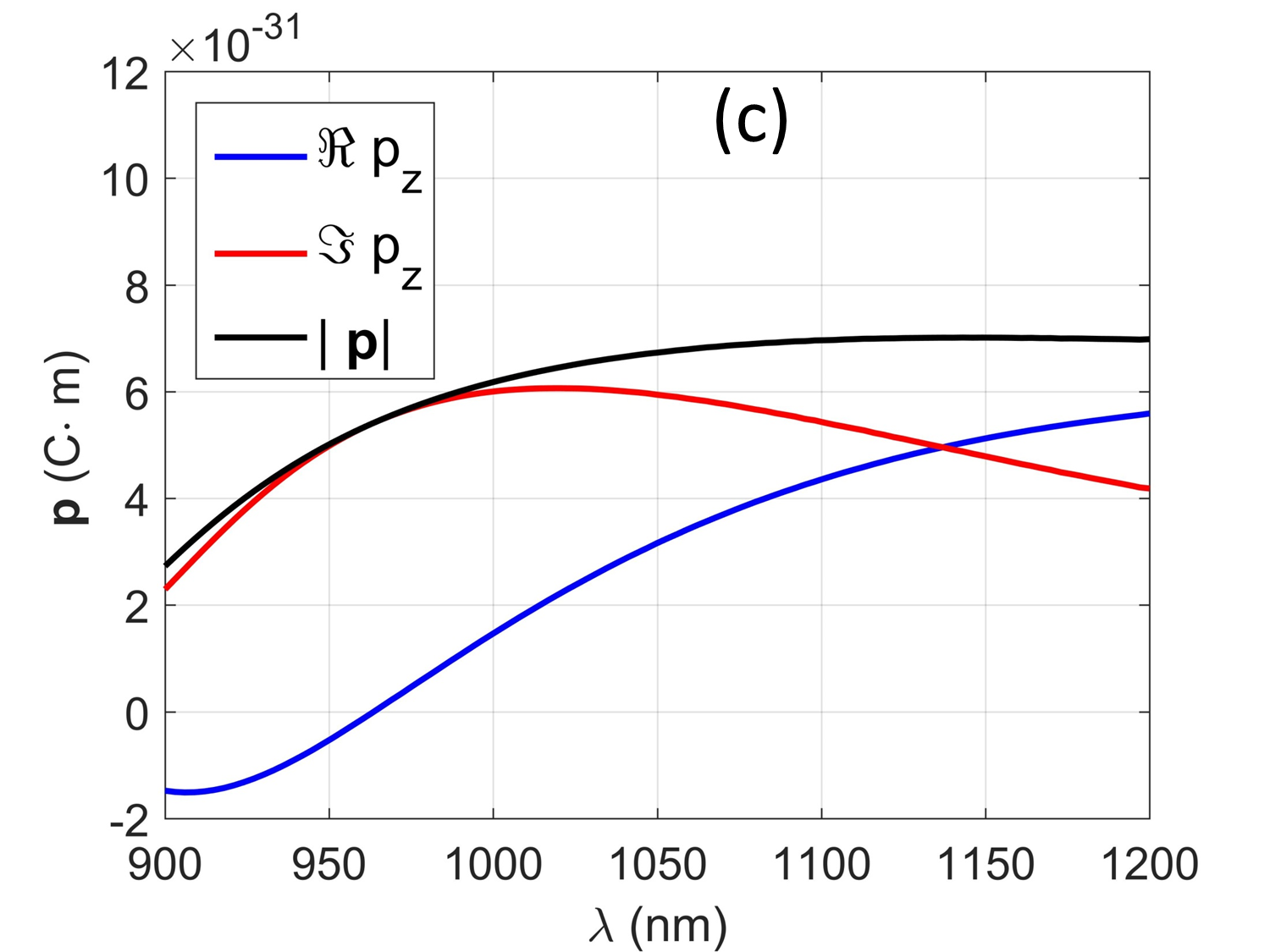}
\includegraphics[width=0.45\linewidth]{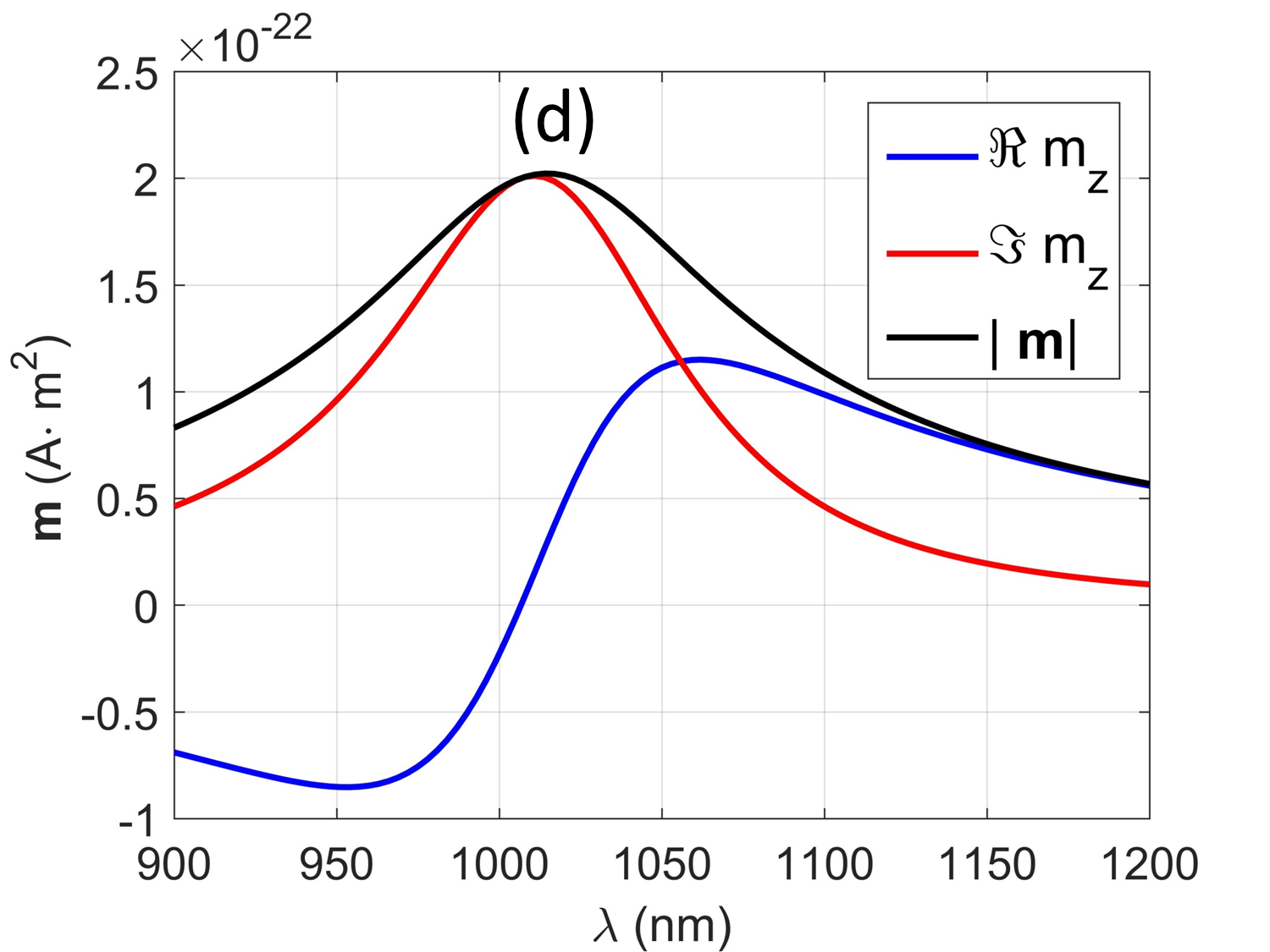}
\caption{(a) and (b) Spectra of the normalized scattering cross sections $\sigma/B$ (where $B$ is the area of the prism base) with corresponding multipole decomposition calculated for the silicon prism. Irradiation conditions are shown in the insets. (c) Components of the electric  dipole moment excited in the prism at the irradiation condition shown in the inset in (a).  (d) Components of the magnetic  dipole moment excited in the prism at the irradiation condition shown in the inset in (b).}
\label{fig4}
\end{figure}

Figures \ref{fig4}a  and  \ref{fig4}b show that for the side irradiation the SCSs are also  determined only by the ED and MD contributions.  In addition, the $E_z$-polarized wave generates the  ED ${\bf p}=(0,0,p_z)$ (see Fig. \ref{fig4}c) and the $H_z$-polarized wave generates the MD  ${\bf m}=(0,0,m_z)$ (see Fig. \ref{fig4}d). Applying the approach described in Ref. \cite{evlyukhin2020bianisotropy}, the polarizabilities $\alpha_{zz}^{e}$ and $\alpha_{zz}^{m}$ can be obtained from equations
\begin{equation}
   \alpha_{zz}^{e}=\frac{p_z^++p_z^-}{2\varepsilon_0 E(0)}, 
\end{equation}
and 
\begin{equation}
   \alpha_{zz}^{m}=\frac{m_z^+-m_z^-}{2(\varepsilon_0/\mu_0)^{1/2} E(0)}, 
\end{equation}
where $\mu_0$ is the vacuum magnetic permeability,  $p_z^+$ and $p_z^-$ are the ED components obtained for the direct (along the $x$-axis) and inverse  irradiation shown in  the inset in Fig. \ref{fig4}a (the electric polarization is the same for both incident wave directions), respectively;  $m_z^+$ and $m_z^-$ are the MD components obtained for the direct (along the $x$-axis) and inverse  irradiation  shown in the inset in Fig. \ref{fig4}b (again the electric polarization is the same  for both incident wave directions), respectively; $E(0)$ is the incident electric field at the point  of the ED and MD location (in this paper this is the electric field amplitude of the incident wave). 

\begin{figure}[htbp]
\centering
%\fbox{\includegraphics[width=\linewidth]{D_6.eps}}
\includegraphics[width=0.8\linewidth]{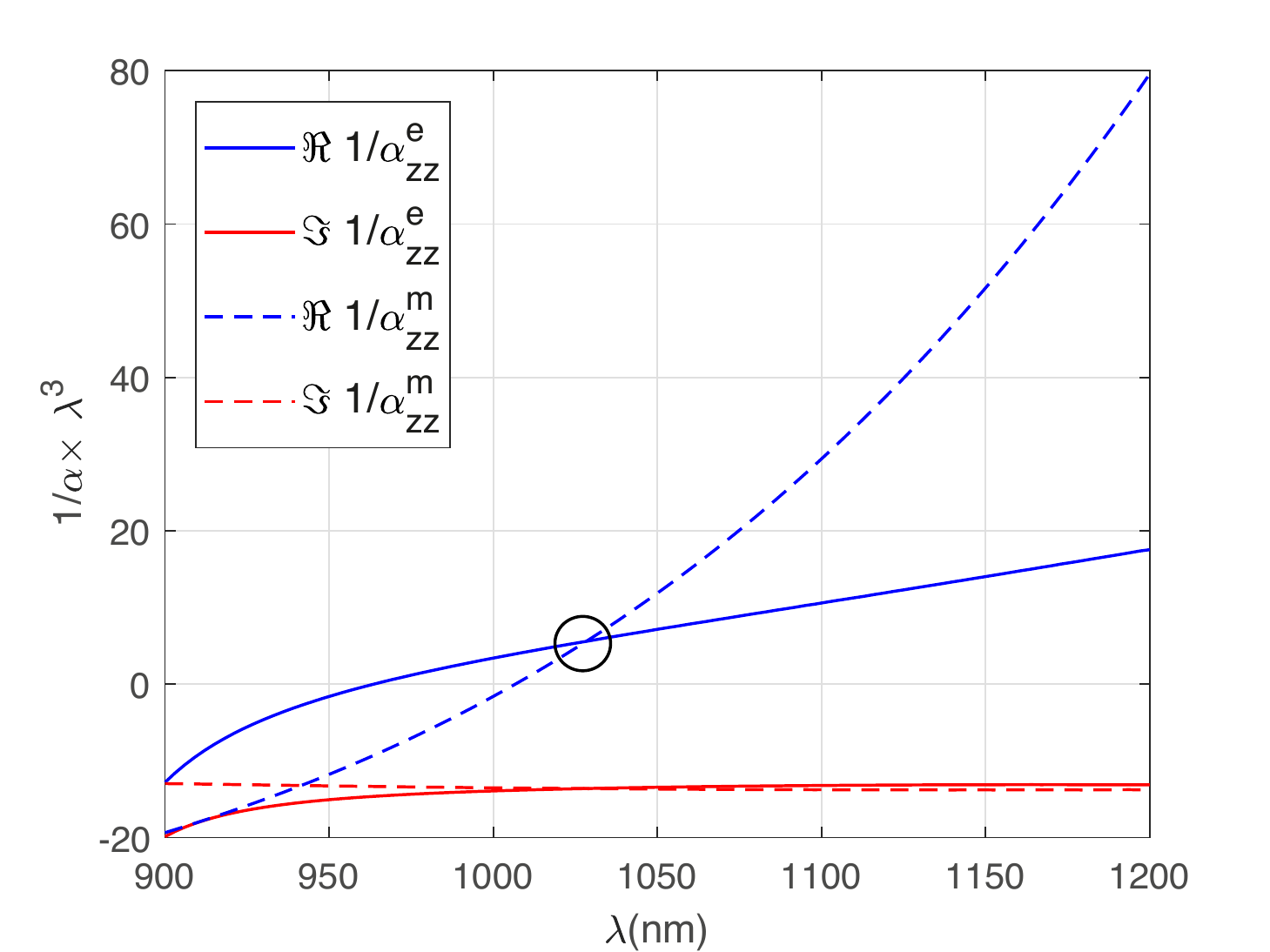}
\caption{Normalized inverse electric and magnetic $zz$-polarizabilities of the prism calculated using the dipole moments presented in Fig. \ref{fig4}c and Fig. \ref{fig4}d, respectively. 
The black circle indicates the point where $\Re(1/\alpha_{zz}^e)=\Re(1/\alpha_{zz}^m)$. }
\label{fig5}
\end{figure}

Spectral behavior  of the normalized inverse polarizabilities is presented in Fig. \ref{fig5}. The imaginary parts of  $\lambda^3/\alpha_{zz}^e$ and $\lambda^3/\alpha_{zz}^m$ have the numerical value close to $(-4\pi^2/3)$ corresponding to Eqs. (\ref{e11}) and (\ref{e12}) with small contributions of bianisotropic (non-local) terms.
The real parts of the normalized inverse polarizabilities intersect at the wavelength $\lambda'\approx1025$ nm and take the  value approximately equal to 5 (see  black circle in Fig. \ref{fig5}). 

To estimate the period $d'$ of the metasurfaces composed of such prisms and supporting the electric and magnetic quasi-trapped modes at the same wavelength, we can apply the (i) procedure described above. Using Fig. \ref{fig1}, we can find that $S'_z\times\lambda^3=5$ at $d/\lambda\approx0.8$ corresponding to $d'=0.8\lambda'=820$ nm, where $\lambda'$ corresponds to the crossing of $\Re(1/\alpha_{zz}^m)$ and $\Re(1/\alpha_{zz}^e)$ in Fig. \ref{fig5}. The next section will show that the obtained period is in complete agreement with the results obtained by more general full-wave numerical simulations.

\section{ Quasi-trapped modes in metasurfaces composed of the prisms  }
 
The dipole method described above gives us information about the periodicity of the metasurface. However, since this is only an approximate approach, in general cases, the metasurface period may differ slightly from that obtained in the dipole approximation for bianisotropic particles. Therefore, the last stage of the proposed procedure should include numerical verification and fine-tuning of metasurface periodicity supporting polarization switching between the quasi-trapped modes.

\begin{figure}[htbp]
\centering
%\fbox{\includegraphics[width=\linewidth]{D_6.eps}}
(a)\includegraphics[width=0.8\linewidth]{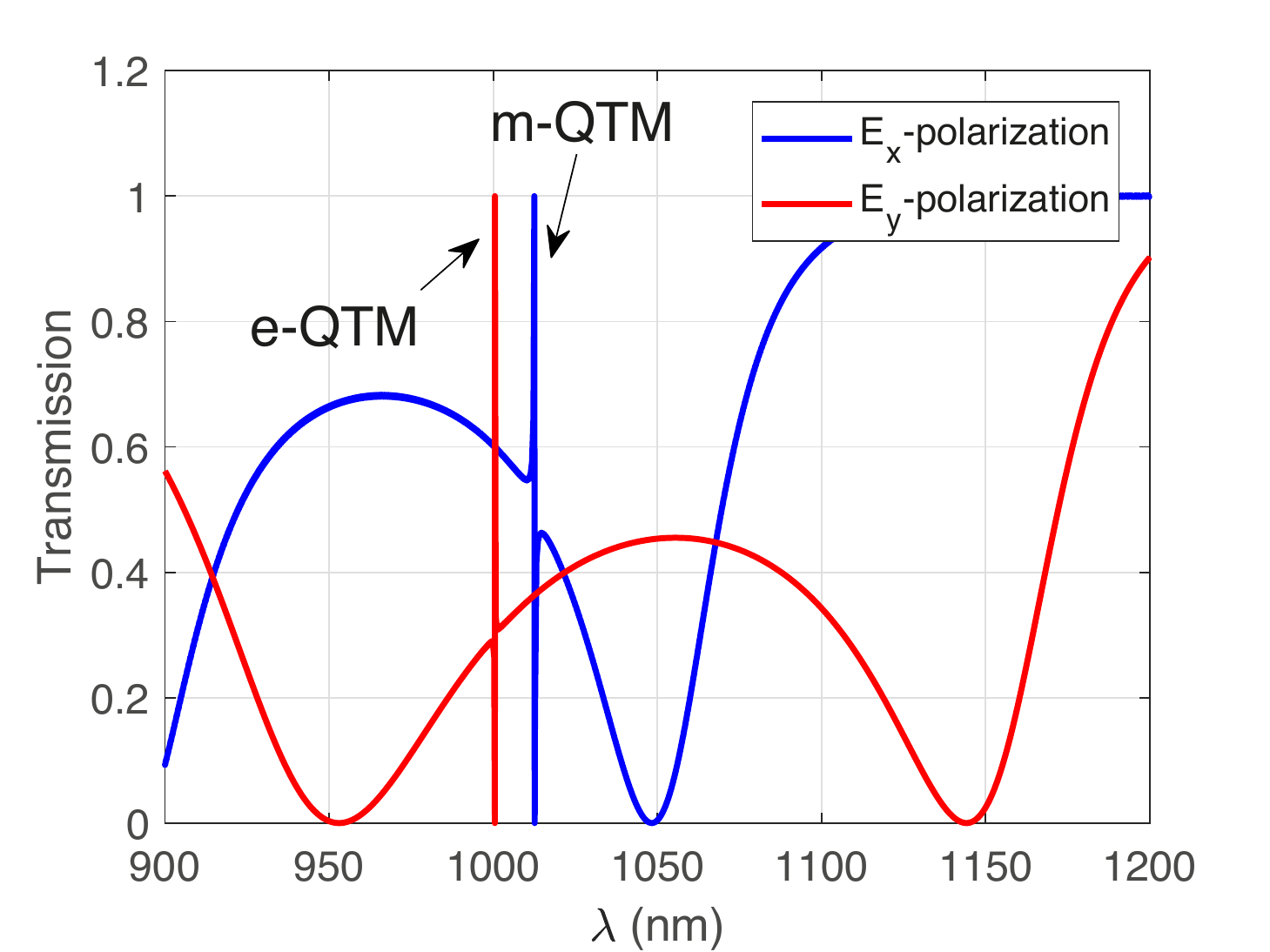}

(b)\includegraphics[width=0.8\linewidth]{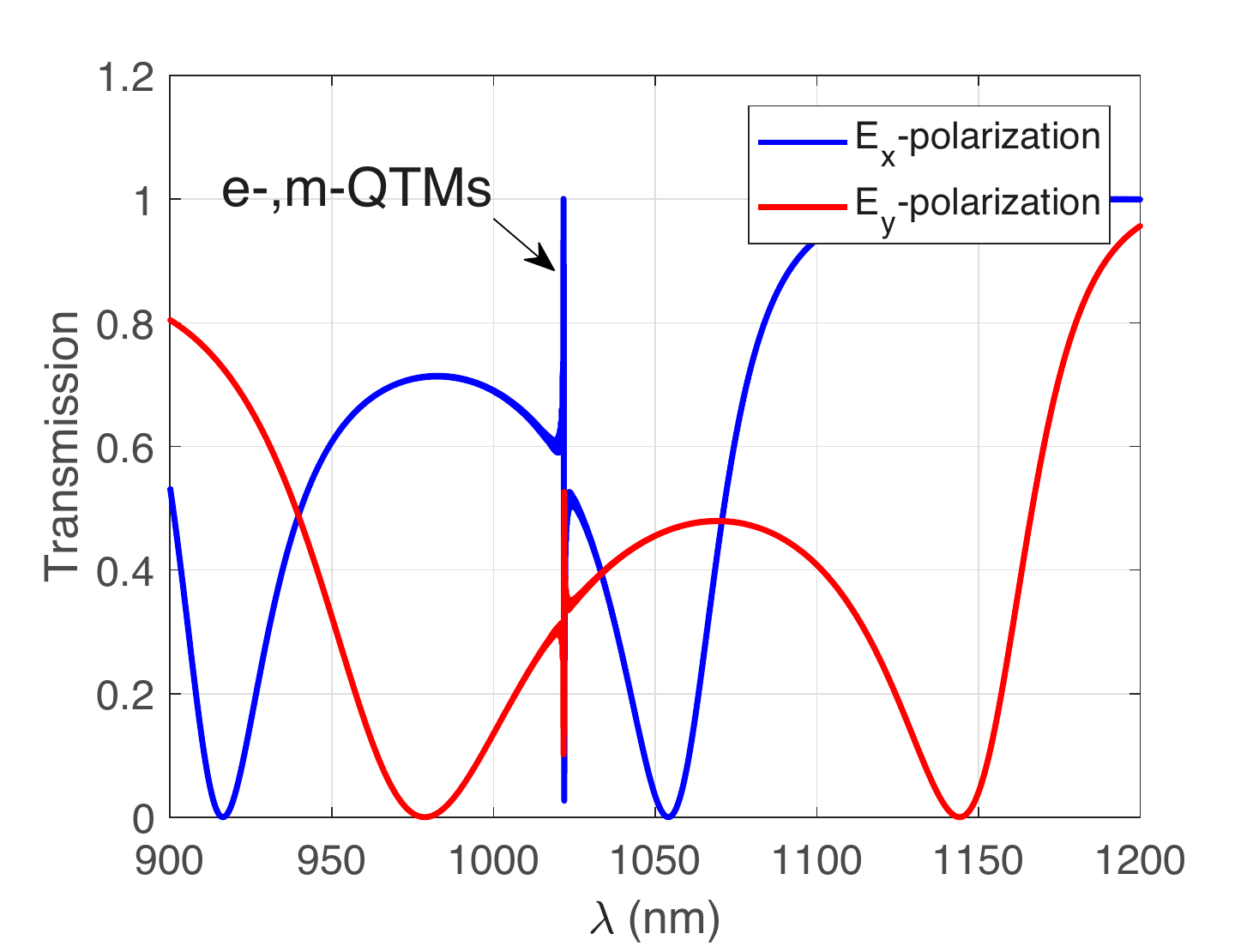}
\caption{Numerically calculated transmission spectra of the metasurfaces composed of the silicon prisms irradiated by normally incident  light plane  waves with different polarizations. The metasurface periods  are  (a) $d=777$ nm and (b) $d=820$ nm.The arrows indicate the electric (e-QTM) and magnetic (m-QTM)  quasi trapped modes (QTMs).  }
\label{fig6}
\end{figure}

Using the Lumerical calculation facilities, we simulate the transmission coefficients of metasurfaces composed of the prisms with parameters corresponding to Fig. \ref{fig4} and orientation as in Fig  \ref{fig2}c with angle $\beta$ directed along $y-$axis (see Fig. \ref{fig3}).  The centers of mass of every prism  are located in the $z=0$ plane. The metasurface is irradiated by normally incident light plane  waves with the electric polarization directed along the $x$- or $y$-axis.  The obtained results are shown in Fig. \ref{fig6}. One can see that for both polarizations, the transmission spectra have narrow resonant features around $\lambda=1000$ nm. For the period $d=777$ nm, these resonances are spectrally separated. However, by tuning the periodicity, it is possible to keep these resonances at the same spectral positions, as shown in Fig. \ref{fig6}b for the period $d=820$ nm. Note that this period $d=820$ nm coincides with the value $d'=820$ nm obtained above in the dipole approximation. 

\begin{figure}[htbp]
\centering
%\fbox{\includegraphics[width=\linewidth]{D_6.eps}}
\includegraphics[width=0.9\linewidth]{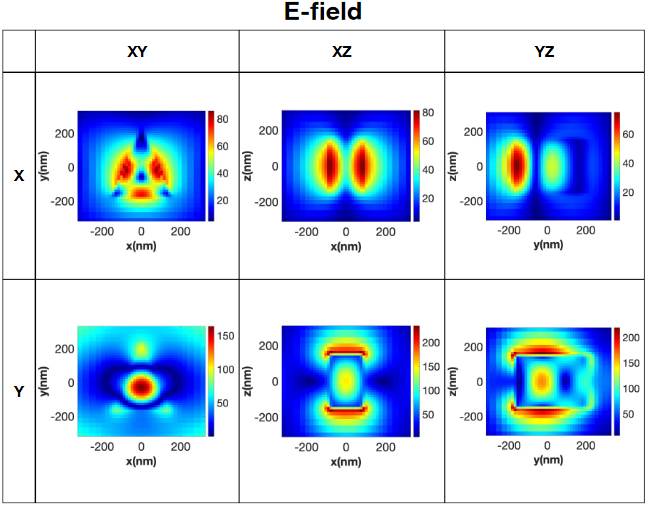}

\includegraphics[width=0.9\linewidth]{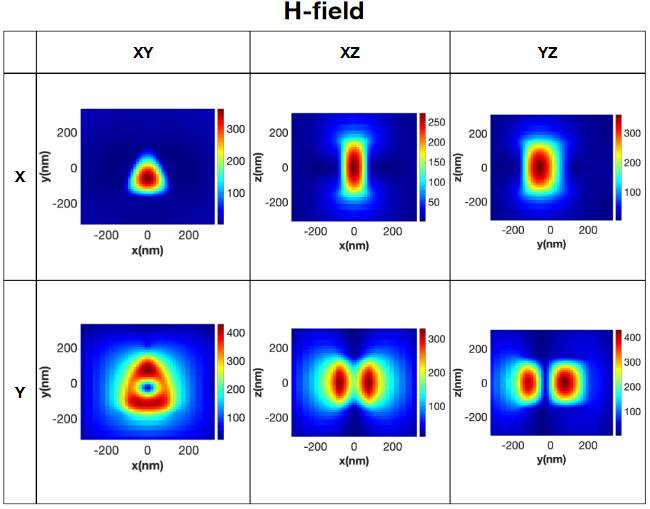}
\caption{Magnitudes of the electric and magnetic field distributions in different planes crossing the metasurface elementary cell containing the prism,  which are calculated at the position of narrow resonances shown in Fig. \ref{fig6}b. The top horizontal lines of the tables define the crossing planes. The vertical left-hand sideline defines the incident wave electric polarization. The magnitudes of the total electric and magnetic fields are normalized  by the amplitudes of the incident wave's electric and magnetic fields, respectively.  }
\label{fig7}
\end{figure}

To prove that these resonances indeed correspond to the electric or magnetic dipole coupling, we present electric and magnetic field distributions in different planes crossing the  elementary cell of the metasurface (see Fig. \ref{fig7}), which are calculated at the position of narrow resonances shown in Fig. \ref{fig6}b. In Fig. \ref{fig7} one can see (for the $xy$-plane) that for the $E_y$-polarization the total electric field is concentrated at the center of the prism,  whereas the magnetic field is distributed like a ring around the electric field. 
Such distribution exactly corresponds to the ED response of the prism. For the  $E_x$-polarization, one has an inverse situation: in the $xy$-plane the magnetic field is concentrated at the prism center, and the electric field has a ring distribution around the center corresponding to the MD response. Thus, at the same spectral position, polarization switching allows excitation of two  resonances corresponding  to  the electric quasi-trapped mode (e-QTM) and  magnetic quasi-trapped mode (m-QTM). One can see in Fig. \ref{fig7} that these modes are associated with the generation of strong electric and magnetic fields inside and around the prisms. Under the e-QTM (m-QTM) excitation, the total electric (magnetic) field has  the maximum value about of 200 (300) times greater than the electric (magnetic) field of the incident wave.   The spatial distribution of these fields can be controlled by the incident waves' polarization (see Fig. \ref{fig7}).  

It is important to emphasize that the metasurface period supporting the  ED and MD quasi-trapped modes at a given spectral position has been determined  in advance by using the polarizabilities of single prisms and the general curve shown in Fig. \ref{fig1}.

\section{Conclusion}
Theoretical approach for the design of metasurfaces, supporting the electric and magnetic quasi-trapped modes (quasi-BICs) located at the same spectral position,  has been developed and demonstrated.  In the framework of dipole approximation, the general conditions for the realization of such modes, connecting single particles’ optical properties with the metasurface lattice periodicity, have been formulated.  Possibilities for tuning these modes to the required spectral position using a combination of analytical and numericalmethods have been suggested. 
This approach has been applied for the development of metasurfaces composed of silicon triangle prisms. It has been shown that one can perform switching between the electric and magnetic quasi-trapped modes located at the same infrared spectral position by changing the incident light polarization. Since the trapped modes are associated with the excitation of strong electric and magnetic fields in the metasurface, the polarization switching opens a way to sub-wavelength manipulation of the electromagnetic fields. This effect creates new opportunities for the application of metasurfaces  with the trapped modes in a wide electromagnetic spectral range from the optical to radio frequencies. 

% Acknowledgements
\medskip
\textbf{Acknowledgements} \par %delete if not applicable))
 The work is supported by the Deutsche Forschungsgemeinschaft (DFG, German Research Foundation) under Germany’s Excellence Strategy within the Cluster of Excellence PhoenixD (EXC 2122, Project ID
390833453) and the Cluster of Excellence QuantumFrontiers
(EXC 2123, Project ID 390837967).  The  models of  the metasurfaces composed of silicon prisms with bianisotropic responses have been supported by the Russian Science Foundation Grant No. 20-12-00343.

% References
\medskip

% Use the following code if you wish to generate your bibliography with BibTeX;
% replace the string "MSP-template" below with the name(s) of
% the BibTeX data base(s) you want to use.
% The resulting bibliography-output (the content of the .bbl file)
% must be pasted back into this file before submission.
% Please also include your BibTeX data base file(s) in your submission
% so that we can re-run BibTeX if necessary.
%
\bibliography{Prism}% Produces the bibliography via BibTeX.

\end{document}